# Developing a General Personal Tutor for Education

Jaan Aru  & Kristjan-Julius Laak

Institute of Computer Science, University of Tartu, Estonia

Correspondence: jaan.aru@gmail.com; julius.laak@gmail.com

## Abstract

The vision of a universal artificial intelligence (AI) tutor has remained elusive, despite decades of effort. Could LLMs be the game-changer? We overview novel issues arising from developing a nationwide AI tutor. We highlight the practical questions that point to specific gaps in our scientific understanding of the learning process.

## The General Personal Tutor

Adapting instruction to meet individual learners' needs through artificial intelligence (AI) has been a quest since the 1950s. Earlier generations of AI tutors, however, were hindered by a range of conceptual issues limiting their use in education [1]. The emergence of large language models (LLMs) has reignited the vision for the General Personal Tutor that delivers high-quality learning experiences across subjects and school levels.

In reality, however, this will not be as simple. LLMs are engineered for frictionless and user-friendly task completion, not for the friction-filled process of learning. Mounting empirical evidence suggests that standard LLM tools (like ChatGPT) do not enhance learning but offload students' thinking, reduce mental effort, fuel metacognitive laziness, and lead to an illusion of learning [2], [3].

Could LLMs be designed to be General Personal Tutors that support the goals of modern education? With its new educational programme, AI Leap, Estonia is taking this question head-on (Box 1). The initiative seeks to reshape the nation's education system so that AI empowers, rather than becomes a barrier to learning. A part of the program is to provide every high school student with access to an AI tutor. However, developing such a General Personal Tutor is not a technological challenge. Rather, the key factors that will determine whether students benefit from such a tutor depend on our understanding of the complexity of learning, pedagogy, and psychology.

The AI Leap's aim to practically implement such a tutor has raised several fundamental questions about how to ground its design in learning sciences. More generally, this work has illuminated the areas where our scientific knowledge of the learning process remains incomplete. Here, we aim to highlight specific issues that have surfaced during the nationwide deployment of the General Personal Tutor.

## Developing the General Personal Tutor

**Which learning sciences' principles should guide AI tutor design?**

To answer that, we could first learn from existing work on training LLMs for educational use. To the best of our knowledge, the only major LLM that is specifically trained for learning is LearnLM by Google DeepMind [6]. In their initial report, they point to the challenge of finding universal pedagogic principles [7]. The challenge lies in the fragmented set of more than thirty (if not hundreds of) learning theories and pedagogic principles, pointing to the absence of a unified theory of classroom learning [8].

The LearnLM team chose five principles to guide the training and evaluation of their model: managing students' cognitive load, inspiring active learning, deepening metacognition, stimulating curiosity, and adapting to students' needs and goals [7]. From the work, it was unclear how those principles were selected. However, this is not a criticism. The more or less holistic understanding of learning in the classroom has not yet translated into an applicable, integrated model of learning.

Similarly, the AI Leap relied on educational expertise to identify key constructs for a General Personal Tutor: to foster autonomous motivation and a growth mindset; scaffold self-regulated learning and metacognition; cue effective learning strategies; and promote conceptual change, all while remaining aligned with curricular objectives. These design choices drew on decades of research in educational and developmental psychology, yet they also underscored a translation gap: despite a rich body of theory and evidence in these research fields, as far as we know, science-backed guidance at the granularity necessary for developing an LLM-based General Personal Tutor is lacking. This is not a critique of prior work. Rather, the gap largely reflects differences in incentives and units of analysis: academic research is rewarded for theory building and studies at the construct or lesson level, but these models need guidance on turn-by-turn, dialogue-level decisions (Figure 1B). Developing AI tutors, therefore, requires translating constructs and principles into concrete model behaviours and assessment routines.

**Which student needs should the system consider?**

Choosing the appropriate approach to support learning in each situation depends on a complex set of students' needs: their motivational, emotional, and cognitive state, as well as their subject-related pre-knowledge and executive functioning. Technically, tracing those needs requires student modelling, a standard technique in adaptive learning systems [10]. However, student models often oversimplify learner needs to constructs such as the current knowledge and skills that are easier to measure by test scores [1]. Combining different theories and principles points to a wider variety of needs that ought to be considered to effectively guide students in their learning journeys.

In other words, a practical AI tutor would need to detect and track students' evolving psychological needs. Here, the Estonian AI Leap team again relied on their academic expertise to compose the list of student needs an AI tutor should ideally include. Some of the needs include, but are not limited to, detecting learners' prior knowledge, illuminating misconceptions and supporting conceptual change [11]; increasing adaptive emotion regulation [12]; managing learning-related beliefs and developing self-regulated learning skills [4]; and supporting autonomous motivation [13]. Although incorporating higher-order goals like metacognitive development into subject-specific adaptive learning systems has been studied for decades [14], LLM tutors like LearnLM do not detect and maintain even simpler student needs. So far, the AI Leap team has used system prompting to steer the existing models towards student need management. However, the main challenge is not only how to compose and implement such a meaningful set of student needs but also how to model their interaction in practical settings.

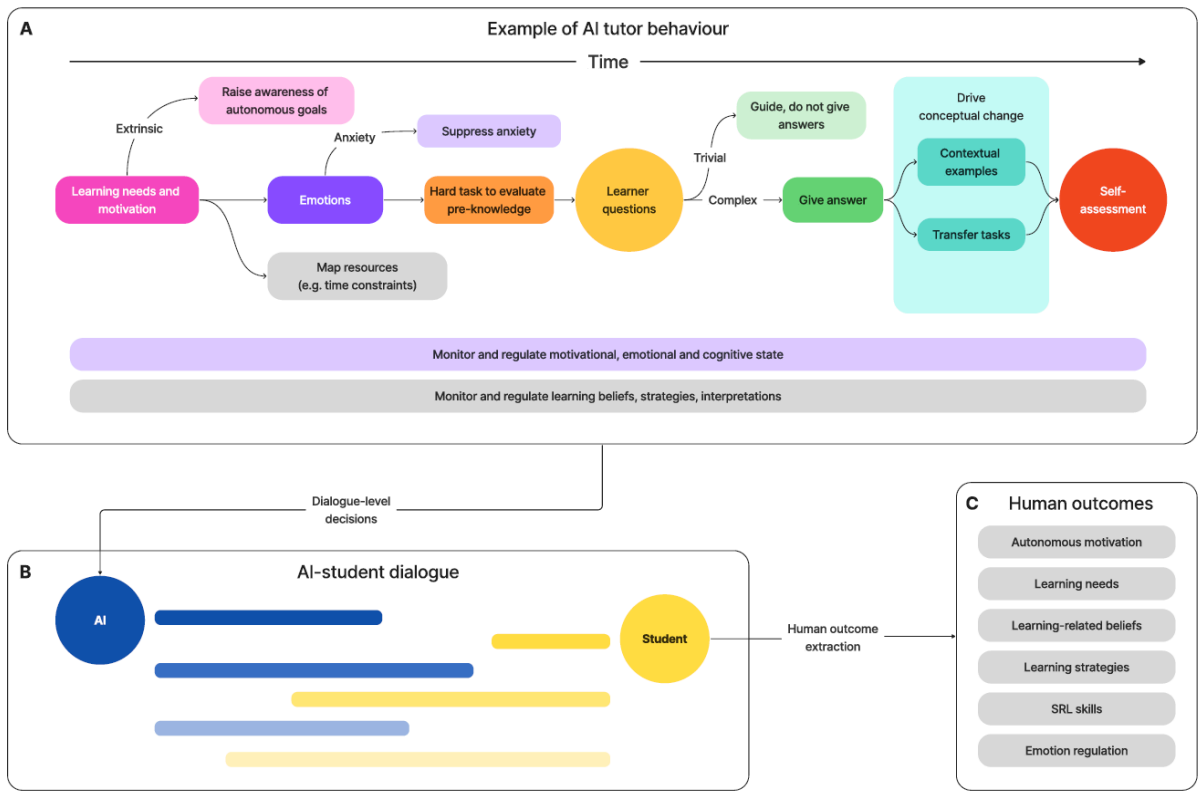

**Figure 1. The complexities of developing an AI tutor for education**
(**A**) Example of how the effectiveness of an AI tutor's behaviour depends on student learning needs, motivation, emotional state, prior knowledge, and pedagogic approaches such as providing answers, promoting conceptual change, and supporting metacognitive processes like self-assessment. (**B**) Our understanding of what is learning guides the AI tutor's behaviour at the level of AI-student dialogue. (**C**) The "success" of the AI tutor is defined by positive changes in human outcomes extracted from the behaviour students exhibit in the chat with the AI tutor.

**How do students' needs influence the learning process?**
Further complicating the issue of student needs is that they are dynamically interlinked to the high-dimensional space of pedagogic principles, resulting in a combinatory explosion in context-dependent pedagogic strategies [7]. For example, in relation to learning-related emotions, an effective AI tutor should, prior to task initiation, transmit positive activating emotions like enjoyment and reduce negative deactivating emotions like anxiety [15], as well as continuously support epistemic emotions like surprise, curiosity and confusion to facilitate further exploration [12] (Figure 1A). Then, students could interact with the tutor in any psychological state and phase of learning (e.g., in the middle of a two-month project). Hence, it is not only about successfully training LLMs to follow the right principles and theories of learning. It is about applying the right technique in the right dosage at the right time [8]. This has remained a challenge for the AI Leap team as there is little previous knowledge of how to incorporate these constructs into interactions with the AI tutor (Figure 1B).

## Success metrics

As it has become clear, a modern General Personal Tutor needs to do more than facilitate domain-knowledge acquisition. The leap we need is not about better test scores. AI Leap is "successful" not when the right principles of learning sciences are trained into the AI tutor [6], but only if, following it, students using it show a sustained positive effect on learning-related beliefs, attitudes, skills, and key educational outcomes (Figure 1C). Without a longitudinal positive effect on those outcomes, there is little benefit from even the most perfect AI tutor.

However, studying the effect of an AI tutor on student outcomes on a national scale raises new challenges. For example, most of the existing measurements require the student to fill in a long questionnaire before and after the intervention, but in practical settings, we would want to continuously detect the change from student-AI dialogues. This would allow researchers to refine the underlying model in response to observed outcomes. Hence, in addition to applying traditional measurements, the Estonian AI Leap aspires to extract those complex human outcomes from the behaviours the students exhibit in the chat. To do that, the team aims to use student-AI dialogues to train machine learning classifiers that would detect constructs like metacognitive skills or cognitive engagement directly from chat content.

## Concluding remarks

The zoo of educational theories and principles of learning illustrates gaps in our understanding of learning, suggesting a need for a framework that would incorporate the complex set of learners' psychological needs and pedagogic processes around them. The rapid advances in AI exert additional pressure.

In summary, applying AI tutors in the education system is not straightforward. Schools and governments should not be swayed by techno-optimism but rather decide for themselves what the goal is. The goal most likely is not to simply bring AI tutors into the classrooms but to first agree on what the learning process is, and based on that knowledge, how to best enhance students' well-being and learning. Research demonstrates that this goal is not achieved by default as a by-product of incorporating vanilla LLMs in schools. Thus, the education systems across the world should either force the technology companies to deliver the tools that meet their goals or take it into their own hands. The success of AI Leap will ultimately depend on how clearly it can outline the ways in which students might gain the most from learning with AI—if such opportunities exist.

**Box 1: Estonian AI Leap blends AI and education**
Estonian AI Leap (*TI-Hüpe*) is a national public–private education initiative, modelled after the landmark "Tiger Leap" of the late 1990s, which provided internet to all Estonian schools, and is set to launch in autumn 2025. Its first phase will provide approximately 20,000 grade 10–11 students and 4,700 teachers with free access to leading AI learning applications, along with comprehensive teacher training.

The program's central technological component is a bespoke AI tutor, developed in collaboration with top AI model developers and grounded in pedagogic principles and scientific theories. For example, the design encourages students to monitor their own understanding, practice retrieval, and persist through challenges, which constitute core aspects of durable learning. Drawing on research in educational psychology, the tutor is intended to act as a scaffold that helps learners gradually internalise effective learning strategies.

Complementing the app, AI Leap includes a flexible, multi-year teacher support program. This features an introductory training module, ongoing professional learning communities, and access to shared materials. The program emphasises co-creation, adaptation based on teacher feedback, and a critical understanding of AI's role in learning. It is worth noting that Estonian teachers have flexibility and autonomy when it comes to implementing how they teach. Hence, not every teacher will apply the AI systems in their teaching. Furthermore, it is up to the teachers whether they let students use these AI tutors only at home or also integrate their usage into classroom practice. Also, nothing prohibits the students from using off-the-shelf AI tools (such as the regular ChatGPT). This all makes conducting scientific studies difficult, as all of these confounding variables need to be measured. And while the AI Leap is not a scientific project, we can only understand its effect on Estonian education by scientifically assessing its progress and consequences.

**Acknowledgements**
We are grateful to Grete Arro for her comments. This work was supported by the Estonian Research Council grants PSG728 and Tem-TA 120, and the Estonian Centre of Excellence in Artificial Intelligence (EXAI), funded by the Estonian Ministry of Education and Research.

**Declaration of interests**
The authors declare no competing interests